\documentstyle[pre,aps,epsfig]{revtex}
\begin{document}
\draft

\title{A Configuration Interaction approach to hole pairing in  \\
the Two-Dimensional Hubbard Model}
\author{E. Louis}
\address{Departamento de F{\'\i}sica Aplicada,
Universidad de Alicante, Apartado 99, E-03080 Alicante, Spain.}
\author{F. Guinea, M. P. L\'opez Sancho, J. A. Verg\'es}
\address{{I}nstituto de Ciencia de Materiales de Madrid,
CSIC, Cantoblanco, E-28049 Madrid, Spain.}
\date{\today}

\maketitle

\begin{abstract}
The interactions between holes in the Hubbard model, in the low
density, intermediate to strong coupling limit, are investigated
by systematically improving mean field calculations. 
The Configuration Interaction basis set is constructed by
applying to  local Unrestricted Hartree-Fock configurations all lattice
translations and rotations. 
It is shown that this
technique reproduces, correctly, the properties of the
Heisenberg model, in the limit of large U.
Upon doping, dressed spin polarons in neighboring sites 
have an increased kinetic
energy and an enhanced hopping rate. Both effects are of the
order of the hopping integral and lead to an effective attraction
at intermediate couplings. 
The numerical results also show that when more than two holes are 
added to the system, they do not tend
to cluster, but rather hole pairs remain far appart. 
Hole--hole correlations are also calculated
and shown to be in qualitative agreement with exact calculations for
$4 \times 4$ clusters.
In particular our results indicate that for intermediate coupling 
the hole--hole correlation attains a maximum when the holes
are in the same sublattice at a distance of $\sqrt{2}$ times
the lattice spacing, in agreement with exact results and the $t$--$J$
model. The method is also used to
derive known properties of the quasiparticle band
structure of isolated spin polarons.
\end{abstract}

\section{Introduction}
The nature of the low energy excitations in the Hubbard model has
attracted a great deal of attention. It is well established that
At half--filling the ground state is an antiferromagnetic (AF) insulator. 
Also, there exists conclusive evidence which indicates that
antiferromagnetism  is rapidly suppressed upon doping \cite{Ka94,Sc95}.
Close to half filling,
a large amount of work suggests the existence of spin polarons,
made of dressed holes, which propagate within a given sublattice
with kinetic energy which in the strong coupling limit is of the order of
$J = \frac{4 t^2}{U}$ \cite{BS94,PL95}, where $t$ is the hopping
integral and $U$ the on site Coulomb repulsion.
These results are consistent with similar calculations
in the strong coupling,
low doping limit of the Hubbard model, the $t-J$ 
model\cite{DN94,LG95,DN97}.
There is also evidence for an effective attraction between these
spin polarons\cite{FO90,BM93,PR94,Da94,KA97,GM98}. However, recent 
and extensive
Monte Carlo calculations for 0.85 filling and $U=2-8t$,
have shown that the pairing correlations vanish as the system size or the
interaction strength increases \cite{ZC97}.

We have recently analyzed the dynamics of spin polarons \cite{LC93a,LC93b}
and the interactions between them \cite{LG98} by means of a systematic 
expansion around mean field calculations of the Hubbard model. 
Two spin polarons in neighboring sites experience an increase
in their internal kinetic energy, due to the overlap of the
charge cloud. This repulsion is of the order of $t$. 
In addition, a polaron reduces the obstacles
for the diffussion of another, leading to an assisted hopping
term which is also of same order. The combination of these
effects is an attractive interaction at intermediate values of
$U/t$. The purpose of this work is to discuss in detail the
results and the approach proposed in \cite{LG98}. We present new results 
which support the validity of our approach, highlighting 
the physically appealing picture of pairing that it provides.
An alternative scheme to go beyond the unrestricted Hartree Fock approximation
is to supplement it with the Gutzwiller projection method, or,
equivalently, slave boson techniques~\cite{SSH98,S98}.
These results are in agreement with the existence of
significant effects due to the delocalization of the solutions,
as reported here.

The rest of the paper is organized as follows. In Section II we discuss
the physical basis of our proposal and the way in which we implement
the Configuration Interaction method. A discussion of the limit of large 
$U/t$ in the undoped case is presented in Section III.
It is shown that, contrary to some expectations,
the Hartree-Fock scheme reproduces correctly the mean field
solution of the Heisenberg model. The systematic
corrections analyzed here can be put in precise correspondence
with similar terms discussed for quantum antiferromagnets.
Results for the $4 \times 4$ cluster
are compared with exact results in Section IV. Section V is devoted to
discuss our results for a single hole (spin polaron) and for two or more
holes. The  hole--hole correlations are also presented
in this Section. The last Section is devoted to the conclusions of
our work. 

\section{Methods}
\subsection{Hamiltonian}
We investigate the simplest version of the Hubbard Hamiltonian
used to describe the dynamics of electrons in CuO$_2$ layers, namely,
\begin{mathletters}
\begin{equation}
H = T + C\;,
\end{equation}
\begin{equation}
T = \sum_{\sigma}T^{\sigma} = -\sum_{\langle ij\rangle}
t_{ij} c^{\dagger}_{i\sigma}c_{j\sigma}\;,
\end{equation}
\begin{equation}
C=\sum_iU_in_{i\uparrow}n_{i\downarrow}\;.
\end{equation}
\end{mathletters}
\noindent The Hamiltonian includes a single atomic orbital
per lattice site with energy $E_i$=0. The sums are over all 
lattice sites $i=1,N_s$ of the chosen cluster of the square lattice
and/or the $z$ component of the spin ($\sigma =\uparrow, \downarrow$).
The operator $c_{j\sigma}$ destroys an 
electron of spin $\sigma$ at site $i$, and $n_{i\sigma}=
c^{\dagger}_{i\sigma}c_{i\sigma}$ is the local density operator.
$t_{ij}$ is the hopping matrix element between sites $i$ and $j$ (the symbol 
$\langle ij\rangle$ denotes that the sum is restricted to all nearest
neighbors pairs) and $U_i$ is the intrasite Coulomb repulsion.
Here we take $t_{ij}=t$ and $U_i = U$, and the lattice constant as
the unit of length.
 
\subsection{Unrestricted Hartree--Fock (UHF) solutions} 
As we shall only consider UHF solutions having a local magnetization
pointing in the same direction everywhere in the cluster, we shall
use the most simple version of the UHF approximation \cite{VL91}.
Within this approximation the effective mean field Hamiltonian
that accounts for the Hubbard term is written as,
\begin{mathletters}
\begin{equation}
C^{\rm eff} = \sum_{\sigma}X^{\sigma} -U\sum_i \langle n_{i\uparrow}
\rangle\langle n_{i\downarrow}\rangle\;,
\end{equation}
\begin{equation}
X^{\sigma}=U\sum_in_{i\sigma}\langle n_{i\sigma}\rangle\;.  
\end{equation}
\end{mathletters}
\noindent The full UHF Hamiltonian is then written as,
\begin{equation}
H^{\rm UHF} = T + C^{\rm eff}\;.
\end{equation}

Use of the Unrestricted Hartree Fock (UHF) approximation in finite clusters
provides a first order approximation to the spin polaron near half
filling. As discussed elsewhere, the UHF approximation
describes well the undoped, insulating
state at half filling \cite{VL91} (see also next Section). 
A realistic picture of the spin wave excitations is obtained by adding
harmonic fluctuations by means of the time dependent Hartree Fock
approximation (RPA)\cite{GL92}. At intermediate and large
values of $U/t$, the most stable HF solution with a single hole is
a spin polaron\cite{VL91,LC93a}. 
In this solution, approximately half of the charge of the hole
is located at a given site.
The spin at that site is small and it is reversed with
respect to the antiferromagnetic background. The remaining charge
is concentrated in the four neighboring sites.
A number of alternative derivations lead to a similar picture
of this small spin bag\cite{Hi87,KS90,DS90,Au94}.
A similar solution is expected to exist in the $t-J$ model.
 
\begin{figure}
\centerline{\epsfig{file=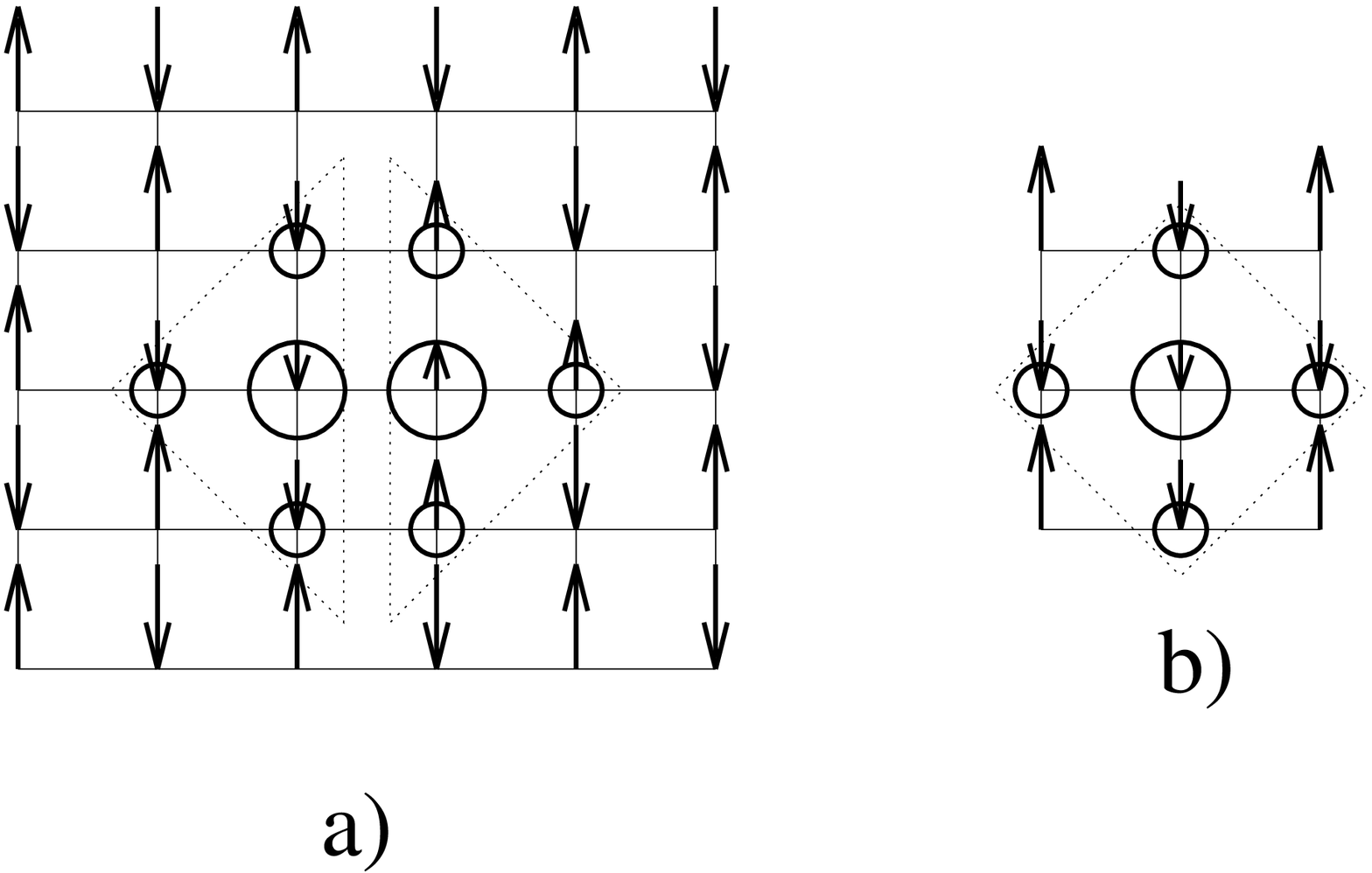,width=3in}}
\caption{
a) Sketch of one of the bipolaron solutions,
at large values of $U/t$, considered in the text.
Circles denote the local charge, measured from half filling,
and arrows denote the spins.
There are two localized states marked by the dashed line.
For comparison, the single polaron solution is shown in b).}
\label{spins}
\end{figure}

A schematic picture of the initial one hole and two holes Hartree Fock
wavefunctions used in this work is shown in Fig. \ref{spins}.
They represent the solutions observed at large
values of $U/t$ for the isolated polaron and two spin polarons
on neighboring sites. The electronic spectrum of these configurations
show localized states which split from the top of the valence band.

As usual in mean field theories, the UHF solutions for an arbitrary
number of holes \cite{VL91}, such as the spin polaron solution described
above, break symmetries which must be restored by quantum fluctuations.
In particular, it breaks spin symmetry and translational
invariance (see Fig. \ref{spins}). 
Spin isotropy must exist in finite clusters. However, it
is spontaneously broken in the thermodynamic limit, due to the
presence of the antiferromagnetic background. Hence, we do not expect
that the lack of spin invariance is a serious drawback of the
Hartree Fock solutions (this point is analyzed,
in some detail in\cite{GL92}). 
Results obtained for small clusters \cite{LC93b,LG92} show a slight
improvement of the energy,
which goes to zero as the cluster size is increased.
On the other hand, translational invariance is expected
to be present in the exact solution of clusters of any size.
The way we restore translational invariance is discussed in the
following subsection. Finally we know how to estimate the effects
due to zero point fluctuations around the UHF ground state \cite{GL92}.
For spin polarons these corrections do not change appreciably
the results, although they are necessary to describe the long
range magnon cloud around the spin polaron \cite{RH97}.

\subsection{Configurations Interaction (CI) method}

We have improved the mean field results by following the procedure
suggested  years ago by some of us \cite{LC93a}.
We hybridize a given spin UHF solution with all
wavefunctions obtained from it by lattice translations.
In the case of two or more holes point symmetry has also to be restored.
This is accomplished by applying rotations to the chosen configuration.
Configurations generated from a given one through this procedure
are degenerate in energy and interact strongly. Here we have also
investigated the effect of extending the basis by including other 
configurations having different energies.
In all cases we include sets of wavefunctions with the lattice symmetry
restored as mentioned. 

In a path integral formulation, this procedure would be equivalent
to calculating the contribution from instantons which visit
different minima. 
On the other hand, it is equivalent to the Configuration Interaction 
(CI) method used in quantum chemistry. 
The CI wavefunction for a solution
corresponding to $N_e$ electrons is then written as
\begin{equation}
\Psi(N_e) = \sum_i a_i \Phi^i(N_e)\;,
\end{equation}
\noindent where the set $\Phi^i(N_e)$ is formed by some chosen UHF 
wavefunctions (Slater determinants) plus those obtained from them by all
lattice translations and rotations. The coefficients $a_i$ are obtained
through diagonalization of the exact Hamiltonian.
The same method, using homogeneous paramagnetic solutions
as starting point, has been used in \cite{FL97}.

The wavefunctions forming this basis set are not in principle
orthogonal. Thus,
both wavefunctions overlap and non--diagonal matrix elements of
the Hamiltonian need to be taken into account when mixing between
configurations is considered. 

If only configurations having the same energy and corresponding, thus,
to the same UHF Hamiltonian, are included, 
a physically sound decomposition of the exact Hamiltonian 
is the following \cite{LC93b},
\begin{equation}
H=H^{\rm UHF} + C -\sum_{\sigma} X^{\sigma} +U\sum_i \langle n_{i\uparrow}
\rangle\langle n_{i\downarrow}\rangle\;,
\end{equation}
\noindent In writing the matrix elements of this Hamiltonian
we should note that the basis formed by the wavefunctions $\Phi_i$
is not orthogonal. Then, we obtain,
\begin{equation}
H_{ij} =\left(E^{\rm UHF}+U\sum_i\langle n_{i\uparrow}
\rangle\langle n_{i\downarrow}\rangle\right)S_{ij}+C_{ij}-
\sum_{\sigma}X_{ij}^{\sigma}S_{ij}^{\bar \sigma}
\end{equation}
\noindent where $E^{\rm UHF}$ is the UHF energy of a given mean field
solution, and the matrix elements of the overlap $S$ are given by
\begin{equation}
S_{ij}=\langle\Phi^i(N_e)|\Phi^j(N_e)\rangle=S_{ij}^
{\uparrow}S_{ij}^{\downarrow}
\end{equation}
This factorization is a consequence of the characteristics 
of the mean field solutions considered in this work (only one
component of the spin different from zero). The specific expression for 
the matrix elements of the overlap is,
\begin{equation}
S_{ij}^{\sigma}=\left |
\begin{array}{ccc}
<\phi_1^{i\sigma}|\phi_1^{j\sigma}> & ... & <\phi_1^{i\sigma}|
\phi_{N_{\sigma}}^{j\sigma}> \\
... & ... & ... \\
<\phi_{N_{\sigma}}^{i\sigma}|\phi_1^{j\sigma}> & ... & 
<\phi_{N_{\sigma}}^{i\sigma}|\phi_{N_{\sigma}}^{j\sigma}>
\end{array}
\right |
\end{equation}
\noindent where the number of particles for each component of the spin are
determined from the usual conditions, $N_{\uparrow}+N_{\downarrow}=N_e$
and $N_{\uparrow}-N_{\downarrow}=2S_z$. The $\phi^{i\sigma}_{n}$
are the monoelectronic wavefunctions corresponding to the Slater
determinant $i$,
\begin{equation}
|\phi^{i\sigma}_{n}\rangle=\sum_k\alpha^{i\sigma}_{nk}c^{\dagger}_
{k\sigma}|0\rangle,
\end{equation}
\noindent $\alpha^{i\sigma}_{nk}$ being real coefficients obtained through
diagonalization of the $H^{\rm UHF}$ Hamiltonian. The matrix element
of the exchange operator between Slater determinants $i$ and $j$ is,
\begin{equation}
X_{ij}^{\sigma}=\left |
\begin{array}{ccc}
<\phi_1^{i\sigma}|X^{\sigma}\phi_1^{j\sigma}> & ... & <\phi_1^{i\sigma}|
\phi_{N_{\sigma}}^{j\sigma}> \\
... & ... & ... \\
<\phi_{N_{\sigma}}^{i\sigma}|X^{\sigma}\phi_1^{j\sigma}> & ... & 
<\phi_{N_{\sigma}}^{i\sigma}|\phi_{N_{\sigma}}^{j\sigma}>
\end{array}
\right |+ ...+\left|
\begin{array}{ccc}
<\phi_1^{i\sigma}|\phi_1^{j\sigma}> & ... & <\phi_1^{i\sigma}|X^{\sigma}
\phi_{N_{\sigma}}^{j\sigma}> \\
... & ... & ... \\
<\phi_{N_{\sigma}}^{i\sigma}|\phi_1^{j\sigma}> & ... &
<\phi_{N_{\sigma}}^{i\sigma}|X^{\sigma}\phi_{N_{\sigma}}^{j\sigma}>
\end{array}
\right |
\end{equation}
\noindent where the matrix elements of $X^{\sigma}$
between monoelectronic wavefunctions are given by,
\begin{equation}
<\phi_n^{i\sigma}|X^{\sigma}\phi_m^{j\sigma}>=U\sum_k
\alpha^{i\sigma}_{nk}\alpha^{j\sigma}_{mk}\langle n_{k{\bar \sigma}}
\rangle
\end{equation}
On the other hand the matrix elements of $C$ are,
\begin{equation}
C_{ij} = U\sum_k \left(n_{k\uparrow}\right)_{ij}
\left(n_{k\downarrow}\right)_{ij} 
\end{equation}
where each $\left(n_{k\sigma}\right)_{ij}$ is given by an equation
similar to Eq. (10). The matrix elements of the density operator
between monoelectronic wavefunctions are,
\begin{equation}
<\phi_n^{i\sigma}|n_{k\sigma}\phi_m^{j\sigma}>=
\alpha^{i\sigma}_{nk}\alpha^{j\sigma}_{mk}
\end{equation}
If the CI basis includes wavefunctions having different UHF energies,
the above procedure is not valid and one should calculate the matrix
elements of the original exact Hamiltonian. Although this is in fact
reduced to calculate the matrix elements of the kinetic energy
operator T, the procedure is slightly more costly (in terms of computer
time) than the one described above. The matrix elements of the
exact Hamiltonian in the basis of Slater determinants are, 
\begin{equation}
H_{ij}=\sum_{\sigma}T_{ij}^{\sigma}S_{ij}^{\bar \sigma}+C_{ij}
\end{equation}
\noindent where $T_{ij}^{\sigma}$ are given by an equation similar 
to Eq. (10), and the matrix elements of the kinetic energy operator
between monoelectronic wavefunctions are
\begin{equation}
<\phi_n^{i\sigma}|T^{\sigma}\phi_m^{j\sigma}\rangle=
-t\sum_{<kl>}\alpha^{i\sigma}_{nk}\alpha^{j\sigma}_{ml}
\end{equation}

The matrix elements involved in the calculation of hole--hole
correlations for a given CI wavefunction are similar to those
that appeared in the computation of the Hubbard term. 
In particular the following expectation value has to be computed,
\begin{equation} 
\langle\Psi|(1-n_k)(1-n_l)|\Psi\rangle=
\sum_{ij}a_ia_j\langle\Phi_i|(1-n_k)(1-n_l)|\Phi_j\rangle
\end{equation}
\noindent where $n_k=n_{k\uparrow}+n_{k\downarrow}$. Terms of four
operators in this expectation value are similar to Eq. (12). Those
requiring more computer time involve
$n_{k\uparrow}n_{l\uparrow}$ with $k\ne l$,
\begin{equation}
\left( n_{k\uparrow}n_{l\uparrow} \right)_{ij} = \left |
\begin{array}{cccc}
<\phi_1^{i\sigma}|n_{k\uparrow}\phi_1^{j\sigma}> & 
<\phi_1^{i\sigma}|n_{l\uparrow}\phi_2^{j\sigma}> &
... & <\phi_1^{i\sigma}|\phi_{N_{\sigma}}^{j\sigma}> \\
... & ... & ... & ... \\
<\phi_{N_{\sigma}}^{i\sigma}|n_{k\uparrow}\phi_1^{j\sigma}> & 
<\phi_{N_{\sigma}}^{i\sigma}|n_{l\uparrow}\phi_2^{j\sigma}> & 
... & <\phi_{N_{\sigma}}^{i\sigma}|\phi_{N_{\sigma}}^{j\sigma}>
\end{array}
\right |+ {\rm permutations}
\end{equation}

\subsection{Numerical Calculations}

Calculations have been carried out on $L \times L$ clusters with
periodic boundary conditions ($L \leq 12$) and $U = 8-5000t$. Some results for
lower values of $U$ are also presented.
Note that $U=8t$ is widely accepted as the most physically meaningful
value of Coulomb repulsion in these systems (see for instance \cite{WS97}).
Although larger clusters can be easily reached, no improvement
of the results is achieved due to the short--range character
of the interactions (see below). 

The numerical procedure runs as follows. Localized UHF solutions are first
obtained and the Slater determinants for a given filling constructed.
The full CI basis set is obtained by applying all lattice traslations
to the chosen localized UHF Slater determinants all having the same $z$
component of the spin $S_z$.
Then we calculate the matrix elements of the overlap and of the Hamiltonian
in that basis set. This is by far the most time consuming part of the
whole calculation. Diagonalization is carried out by means of standard 
subroutines for non--orthogonal bases. The state of lowest energy 
corresponds to the CI ground state of the system for a given $S_z$. 
The desired expectation values are calculated by means of this ground state 
wavefunction. The procedure is variational and, thus, successive 
enlargements of the basis set do always improve the description
of the ground state. 

\section{The limit of large $U$ in the undoped case.}

The Hartree Fock scheme, for the undoped Hubbard model in a square lattice
gives an antiferromagnetic ground state, with a charge gap.
At large values of $U/t$, the gap is of order $U$. The simplest
correction beyond Hartree-Fock, the RPA approximation, leads to
a continuum of spin waves at low energies\cite{GL92}. 
Thus, the qualitative features
of the solution are in good agreement with the expected properties
of an antiferromagnetic insulator. 

There is, however, a great deal
of controversy regarding the adequacy of mean field techniques 
in describing a Mott insulator\cite{La97,AB97}. 
In principle, the Hubbard model, in the large $U$ limit,
should describe well such a system. At half filling and large $U$,
the only low energy degrees of freedom of the Hubbard model 
are the localized spins, which interact antiferromagnetically,
with coupling $J = \frac{4 t^2}{U}$. It has been argued that,
as long range magnetic order is not relevant for the existence
of the Mott insulator, spin systems with a spin gap are the most
generic realization of this phase. A spin gap is often associated
with the formation of an RVB like state,
which cannot be adiabatically connected to the Hartree Fock
solution of the Hubbard model. So far, the best
examples showing these features are two leg spin 1/2 ladders\cite{DR96}.
Recent work\cite{CK98} indicates that, in the presence of
magnetic order, the metal insulator transition is of the Slater
type, that is, coherent quasiparticles can always be defined
in the metallic side. These results seem to favor the scenario
suggested in\cite{La97}, and lend support to our mean field 
plus corrections approach.

Without entering into the full polemic outlined above, we now show
that  the method used here gives, in full detail, the results which
can be obtained from the Heisenberg model by expanding around the
antiferromagnetic mean field solution\cite{CK98}. Such an expansion
gives a consistent picture of the physics of the Heisenberg model
in a square lattice.

The ground state energy of the Hartree Fock solution in a $4 \times 4$
cluster is compared to the exact value\cite{FO90} at large values of $U$
in Table I. The corresponding Heisenberg model is:
\begin{equation}
H_{\rm Heis} = \frac{4 t^2}{U} \sum_{ij} {\bf \vec{S}}_i {\bf \vec{S}}_j
- \frac{t^2}{U} \sum_{ij} n_i n_j
\label{Heisenberg}
\end{equation}
In a $4 \times 4$ cluster, the exact ground state energy is
\begin{equation}
E_{\rm Heis} = - 16 ( c + 0.5 )\frac{4 t^2}{U} 
\end{equation}
\noindent where $c = 0.702$ \cite{Sa97}, in good
agreement with the results for the Hubbard model. The mean field energy
can be parametrized in the same way, except that $c = 0.5$. This is the
result that one expects for the mean field solution of the
Heisenberg model, which is given by a staggered configuration of static
spins. This solution can be viewed as the ground state of an anisotropic
Heisenberg model with $J_z = J$ and $J_{\pm} = 0$. 

We now analyze corrections to the Hartree Fock solution by hybridizing it
with mean field wavefunctions obtained from it by flipping two
neighboring spins (hereafter referred to as sf). These solutions are 
local extrema of the mean field solutions in the large $U$ limit.
In Table I we show the energy difference between
these states and the antiferromagnetic (AF) Hartree Fock ground state, 
and their
overlap and matrix element also with the ground state. We have checked
that these are the only wavefunctions with a non negligible mixing with
the ground state. The overlap goes rapidly to zero, and the energy difference
and matrix elements adjust well to the expressions 
\begin{mathletters}
\begin{equation}
\Delta E_{\rm AF,sf} = E_{\rm AF}-E_{\rm sf} = \frac{12 t^2}{U}
\end{equation}
\begin{equation}
t_{\rm AF,sf} = \frac{2 t^2}{U}\;.
\end{equation}
\end{mathletters}
These are the results that one obtains when proceeding from
the Heisenberg model. These values, inserted in a perturbative analysis of
quantum corrections to the ground state energy of the Heisenberg
model\cite{CK98}, lead to excellent agreement with exact results 
(see also below).

As already pointed out, in the CI calculation of the ground state energy 
we only include the mean field
wavefunctions with two neighboring spins flipped. 
Restoring point symmetry gives
a total of 4 configurations, while applying lattice translations 
leads to a set of $4L^2/2$ configurations (remember that
configurations on different sublattices do not interact) to which
the AF wavefunction  has to be added. In the case
of the $4 \times 4$ cluster the set has a total of 33 configurations. The
CI energy for this cluster is given in Table I along with the exact
and the UHF energies. It is noted that the CI calculation reduces
in 50\% the difference between the exact and the mean field result.
Improving this results would require including a very large set, 
as  other configurations do only decrease the ground state energy 
very slightly.

In the large U limit, the largest interaction is $t_{\rm AF,sf}$. Then,
neglecting the overlap between the AF and the sf mean field solutions,
the CI energy of the ground state  can be approximated by,
\begin{equation}
E_{\rm CI}=\frac{1}{2}\left[ E_{\rm AF}+E_{\rm sf} - \Delta E_{\rm AF,sf}
\sqrt{1+\frac{8L^2t_{\rm AF,sf}^2}{(\Delta E_{\rm AF,sf})^2}}\right]
\end{equation}
\noindent For $U=50$ this expression gives $E_{\rm CI}$=-1.421,  
in excellent agreement with the CI result given in Table I.

Note that a perturbative calculation of the corrections of the 
ground state energy in finite clusters is somewhat tricky, 
as the matrix element
scales with $\sqrt{N_s}$, where $N_s=L^2$ is the number of sites in the cluster,
while the energy difference is independent of $N_s$. The first term
in a perturbative expansion (pe) coincides with the first term in the
expansion of the square root in Eq. (21),
\begin{equation}
E_{\rm pe}=E_{\rm AF}-\frac{2L^2t_{\rm AF,sf}^2}{\Delta E_{\rm AF,sf}}
\end{equation}
\noindent in agreement with the result reported in \cite{CK98}. Although
this correction to the AF energy has the expected size dependence for
an extensive magnitude  (it is
proportional to the number of sites $N_s$) and gives an energy 
already very similar to the exact, it was obtained by inconsistently
expanding in terms of a parameter that can be quite large. For instance
in the $4 \times 4$ cluster and $U$=50, $E_{\rm pe}\approx 1.51$, close
to the exact result (Table I) while 
$(8L^2t_{\rm AF,sf}^2)/(\Delta E_{\rm AF,sf})^2 \approx 3.9$ 
much larger than 1. Thus, perturbation theory
is doomed to fail even for rather small clusters. 

On the other hand, the CI calculation described above introduce a
correction to the AF energy which has not the correct size dependence.
This can be easily checked in the large cluster limit in which
the CI energy can be approximated by
\begin{equation}
E_{\rm CI}\approx \frac{1}{2}\left( E_{\rm AF}+E_{\rm sf}\right) - 
\sqrt{2}L t_{\rm AF,sf}
\end{equation}
\noindent while the correct expression should scale as $N_s$,
because, in large clusters, the difference between the exact and the
Hartree-Fock ground state energies must be proportional to $N_s$, irrespective
of the adequacy of the Hartree-Fock approximation. 

Thus, one obtains a better approximation to the ground state
energy in the thermodynamic limit, by using
the perturbative calculations in small clusters and extrapolating
them to large clusters, as in the related $t-J$ model\cite{CK98}.
In any case, the problem outlined here does not appear when
calculating corrections to localized spin textures, such as the
one and two spin polarons analyzed in the next sections. 
The relevant properties are associated to the size of the texture,
and do not scale with the size of the cluster they are embedded in.

 From the previous analysis, we can safely conclude that our scheme
gives a reliable approximation to the undoped Hubbard model in a square
lattice in the strong coupling regime, $U/t \gg 1$. We cannot conclude
whether the method is adequate or not for the study of models which exhibit
a spin gap. It should be noted, however, that a spin gap needs not
only be related to RVB like ground states. A spin system modelled
by the non linear sigma model can also exhibit a gap in the
ground state, if quantum fluctuations, due to dimensionality or
frustration, are sufficiently large. In this case, a mean field approach
plus leading quantum corrections should be qualitatively correct.

\section{Comparison with Exact Results for $4 \times 4$ clusters 
with two holes}

In order to evaluate the performance of our approach we have calculated
the ground state energy of  two holes in the $4 \times 4$ cluster
and compared the results with those obtained by means of the Lanczos method
\cite{FO90}. The results are reported in Tables II--IV, where the energies
for one hole are also given for the sake of
completness (a full discussion of this case can be found 
in \cite{LC93a}, see also below).
In the case of one hole the standard spin polaron solution (Fig. \ref{spins})
plus those derived from it through lattice translations form the basis
set. For two holes we consider solutions with $S_z=0$ or 1. In the first 
case we include either the
configuration having the two holes at the shortest distance, i.e., 
separated by a (1,0) vector and/or at the largest distance possible, that
is separated by a (2,1) vector, and those obtained from them
through rotations. The basis used for the two polarons at the shortest 
distance is shown in Fig. \ref{bipolaronCI}. The set of these four 
configuration has the proper point symmetry.
Again, lattice translations are applied to these configurations to 
produce the basis set with full translational symmetry. 
On the other hand, wavefunctions with $S_z=1$ can be constructed by including 
configurations with the two holes separated by vectors (1,1) and/or (2,2).

The results for the energies of wavefunctions with $S_z=0$ for several
values of the interaction parameter $U$ are reported in Tables II and III.
As found for a single hole, the kinetic energy included by restoring
the lattice symmetry, improves the wavefunction energies \cite{LC93a}.
The improvement in the energy is larger for intermediate $U$. 
For instance for $U=32$ a 10\% gain is noted.  
Within UHF, the solution with the holes at the largest distance 
is more favorable for $U > 8t$. Instead, restoring the translational
and point symmetries favors the solution with the holes at neighboring
sites for all $U$ shown in the Tables. The results also indicate that
the correction introduced by this procedure does no vanish with $U$.
A more detailed discussion
of the physical basis of this result along with results for larger values
of $U$ and larger clusters will be presented in the following Section. 
On the other
hand the energies get closer to the exact energies (see  Table II).
A further improvement in the energy is obtained by including 
both UHF configurations, namely, \{1,0\} and \{2,1\}. This
improvement is larger for 
intermediate $U$ and vanishes as $U$ increases (Table III). 
Other configurations, such as that proposed in \cite{RD97} in which the two 
holes lie on neighboring sites along a diagonal and a neighboring spin
is flipped, may contribute to further improve the CI energy of
the ground state. 

It is interesting to compare these results with those corresponding
to wavefunctions with $S_z=1$ also reported in Table IV. It is noted that
for $U=6-16$ the energy of the solution including all configurations 
from the set \{1,1\} is smaller than those obtained with all configurations
from either the set \{1,0\} or the set \{2,1\}. However, the wavefunction
constructed with all configurations from the last two sets is
more favorable than the best wavefunction with $S_z=1$. The latter is
in agreement with exact calculations \cite{FO90} which obtained
a ground state wavefunction with $S_z=0$.

\section{Results}

\subsection{Single Polaron}
Here we only consider the quasiparticle band structure associated
to the single polaron, the energy gain induced through  restoration
of translational symmetry has been considered elsewhere \cite{LC93a}. 
The calculated dispersion band of a single polaron is shown in 
Fig. \ref{polaron}.  Because of the antiferromagnetic background,
the band has twice the lattice periodicity.
Exact calculations in finite clusters do not show
this periodicity, as the solutions have a well defined spin
and mix different background textures. As cluster sizes are increased,
however, exact solutions tend to show the extra periodicity
of our results. We interpret this as a manifestation
that spin invariance is broken in the thermodynamic
limit, because of the antiferromagnetic background.
Hence, the lack of this symmetry in our calculations
should not induce spurious effects.
Fig. \ref{polaron} shows the polaron bandwidth
as a function of $U$. It behaves as $t^2/U$, the fitted law being
\begin{equation}
E_{\rm BW}=-0.022 t + 11.11 \frac{t^2}{U}
\end{equation} 
\noindent This result indicates that the band width tends to zero
as $U$ approaches infinite, as observed in the results
for the energy gain reported in \cite{LC93a}.
Our scheme allows a straightforward explanation of
this scaling. Without reversing the spin of the whole background,
the polaron can only hop within a given sublattice.
This implies an intermediate virtual hop into a site with
an almost fully localized electron of the opposite spin.
The amplitude of finding a
reversed spin in this new site decays as $t^2/U$ at large $U$.

On the other hand, we find that the dispersion relation 
can be satisfactorily fitted by the expression:  
\begin{eqnarray}
\epsilon_{\bf k} = \epsilon_0 + 4 t_{11} \cos (k_x ) \cos ( k_y ) 
 + 2 t_{20} [ \cos ( 2 k_x ) + \cos ( 2 k_y ) ] 
+  4t_{22} \cos ( 2 k_x ) \cos ( 2 k_y ) +\nonumber \\
4 t_{31}[ \cos ( 3 k_x) \cos ( k_y )+ \cos ( k_x ) \cos ( 3 k_y )].
\end{eqnarray}
\noindent For $U = 8 t$, we
get $t_{11} = 0.1899 t$ , $t_{20} = 0.0873 t$, $t_{22} = -0.0136 t$,
and $t_{31} = -0.0087 t$. All hopping integrals vanish as $t^2/U$
in the large $U$ limit for the reason given above.
Also the energy gain with respect to UHF \cite{LC93b}
behaves in this way. All
these features  are in good agreement with known
results\cite{BS94,PL95,DN94,LG95} for both
the Hubbard and the $t-J$ models.

\subsection{Two Holes}
\begin{figure}
\centerline{\epsfig{file=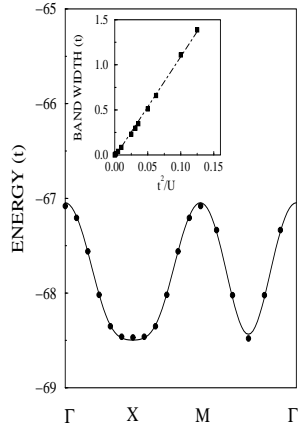,width=3in,height=3in}}
\caption{
Sketch of the bipolaron UHF wavefunctions used in this work. Note that the 
four wavefunctions are obtained by successive rotations
of $\pi /2$. The complete basis set is produced by translation of these
wavefunctions through the whole cluster.}
\label{bipolaronCI}
\end{figure}

\begin{figure}
\centerline{\epsfig{file=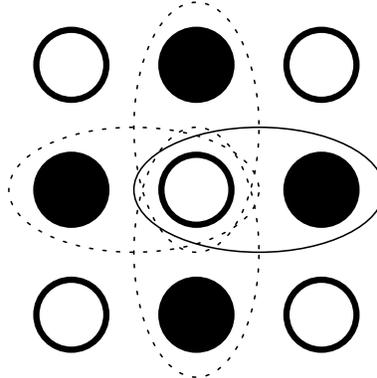,width=2in}}
\caption{
Quasiparticle band structure for a single hole on $12 \times 12$ clusters
of the square lattice with periodic boundary conditions and $U=8$.
The continuous line corresponds to the fitted
dispersion relation (see text). The inset
shows the bandwidth as a function of $t^2/U$ for $U \ge 8t$;
the fitted straight line is $-0.022 t + 11.11 t^2/U$.}
\label{polaron}
\end{figure}

We now consider solutions with two spin polarons.
The relevant UHF solutions are those with $S_z = 0$ (solutions with
$S_z=1$ will also be briefly considered).
In order for the coupling to be finite, the centers of the
two spin polarons must be located in different sublattices.
The mean field energy increases as the two polarons
are brought closer, although, for intermediate and large
values of $U$, a locally stable Hartree Fock solution can be
found with two polarons at arbitrary distances. We have not attempted 
to do a full CI analysis of all possible combinations of two holes in a finite
cluster. Instead, we have chosen a given mean field solution (UHF)
and hybridized it with all others obtained by all lattice translations
and rotations. Some results of calculations in which more than one 
UHF solution is included will be also presented. Clusters of sizes up to 
$10 \times 10$ were studied
which, as in the case of the polaron, are large enough due to the 
short--range interactions between different configurations. 
The basis used for the two polarons at the shortest distance
is shown in Fig. \ref{bipolaronCI}.
This procedure leads to a set of bands, whose number  depends on
the number of  configurations included in the CI calculation.
For instance if the basis set of Fig. \ref{bipolaronCI} is used four 
bands are obtained (see also below). 

\begin{figure}
\centerline{\epsfig{file=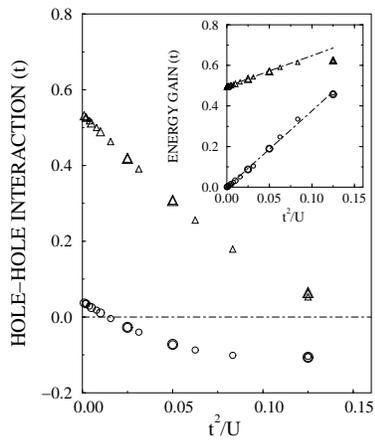,width=3in}}
\caption{
Comparison of the hole--hole interaction (see main text for the definition)
obtained within UHF (triangles) and CI (circles) approximations
for wavefunctions with $S_z=0$. Results correspond to $6 \times 6$, 
$8 \times 8$ and $10 \times 10$ clusters with periodic boundary
conditions, and $U \ge 8t$. The size of the symbols increases with
increasing cluster size.
The inset shows the energy gain due to the inclusion of correlation
effects via CI for both the configuration of holes located in
neighboring positions (circles) and holes that are maximally
separated in the finite size cluster (triangles).
The respective asymptotic behaviors for large $U$ are:
$0.495 t + 1.53 t^2/U$ for holes at the shortest distance and
$-0.002 t + 3.78 t^2/U$ for holes at the largest distance.}
\label{difference}
\end{figure}

Like in the single polaron case,
we obtain a gain in energy (with respect to UHF), due to the delocalization 
of the pair.  The numerical results for $L$=6, 8 and 10 and $U$
in the range $8t-5000t$ are shown in the inset of Fig. \ref{difference}.
They can be fitted by the following straight lines,
\begin{mathletters}
\begin{equation}
E_{\rm G}^{{1,0}}=0.495 t + 1.53 \frac{t^2}{U}
\end{equation}
\begin{equation}
E_{\rm G}=-0.002 t + 3.78 \frac{t^2}{U}
\end{equation}
\end{mathletters}
\noindent where (26a) corresponds to holes at the shortest
distance and (26b) to holes at the largest distance. 
Note that, whereas in the case of the holes at the largest distance, 
the gain goes to zero
in the large $U$ limit, as for the isolated polaron, when
the holes are separated by a $\{1,0\}$ vector the gain goes to a
finite value. This result is not surprising, as the following arguments 
suggest. The hopping terms in the bipolaron calculation,
that are proportional to $t$ at large $U$,
describe the rotation of a pair around the position of one of
the two holes. Each hole is spread between four sites.
In order for a rotation to take place, one hole has to jump
from one of these sites into one of the rotated positions.
This process can always take place without a hole moving into a
fully polarized site with the wrong spin. There is a gain
in energy, even when $U/t \rightarrow \infty$. In the single
polaron case, the motion of a hole involves the inversion of, at least, one 
spin, which is fully polarized in the large $U$ limit. Because of this, 
hybridization gives a vanishing contribution to the energy as $U/t 
\rightarrow \infty$.

The results discussed above are in line with those for the width of the
quasiparticle band. The numerical results can be fitted by,
\begin{mathletters}
\begin{equation}
E_{\rm BW}^{{1,0}}=3.965 t + 14.47 \frac{t^2}{U}
\end{equation}
\begin{equation}
E_{\rm BW}=-0.007 t + 10.1 \frac{t^2}{U}
\end{equation}
\end{mathletters}
Thus, the total bandwidth of the two bands obtained for holes in neighboring
sites does not vanish in the infinite $U$ limit (as the energy gain reported
in Fig. \ref{bipolaronCI}). The internal consistency of
our calculations is shown comparing the large $U$ behavior
of the two holes at the largest distance possible with the corresponding
results obtained for the isolated polaron (compare this fitting with
that given in Eq. (24)) .

The hole--hole interaction, i.e., the difference
between the energy of a state built up by all configurations with the
two holes at the shortest distance (separated by a vector of the set
\{1,0\}) and the energy of the state having the holes at the largest
distance possible at a given cluster is depicted in Fig. \ref{difference}.
Two holes bind for intermediate values of $U$ \cite{ferro}. 
This happens because the delocalization energy tends to be higher
than the repulsive contribution obtained within mean field.
The local character of the interactions is
illustrated by the almost null dependence of the results
shown in Fig. \ref{difference} on the cluster size. 

The only numerical calculations which discuss the binding of holes
in the Hubbard model are those reported in \cite{FO90}. Energetically,
it is favorable to pair electrons in a $4 \times 4$ cluster for
values of $U/t$ greater than 50. The analysis of the correlation functions
suggests that pairing in real space ceases at $U/t \sim 16$,
leading to the suspicion of large finite size effects.
Our results give that pairing dissappears at $U/t \approx 40$,
which is consistent with the analysis in\cite{FO90}.
Similar calculations for the t-J model give binding between
holes for $J/t \ge 0.1$\cite{PR94}. Taking $J = \frac{4 t^2}{U}$,
this threshold agrees well with our results. 

In order to clarify some aspects of the method, we have
carried out a more detailed analysis of two hole solutions in
$6 \times 6$ clusters. The results are presented in Table V.
Within UHF the most favorable solution is that with the two
holes at the largest distance (2,3) but for the smallest $U$
shown in the Table. The solution with the holes at the shortest
distance (1,0) is only favored at small $U$, while for $U\geq 8$ 
even the solution with $S_z=1$ has a smaller energy. 
Instead when the lattice symmetry is restored the solution
with the holes at the shortest distance is the best for all $U$ 
excluding  $U=200$. For
such a large $U$ the wavefunction constructed with all configurations
from \{2,3\} has the lowest energy. The solution with $S_z=1$ is
unfavorable for all $U$ shown in  Table V, in contrast with
the results found in the $4 \times 4$ cluster, indicating that
size effects were determinant in the results for the smaller cluster.
Including all configurations with $S_z$ either 1 or 0 does not change 
this trend. The small difference between the results 
for \{1,0\} and those with
all configurations with $S_z=0$ for large $U$ is misleading. In fact,
the weight of the configuration with the holes at the largest
distance (2,3) in the final CI wavefunction increases with $U$. 
This will be apparent in the hole--hole correlations
discussed in the following paragraph.

We have analysed the symmetry of the ground state wavefunction 
$|\Psi\rangle$ obtained with all configurations having the holes
at the shortest distance. The numerical results for all $U$ show
that $\langle\Phi^1|\Psi\rangle=-\langle\Phi^2|\Psi\rangle=
\langle\Phi^3|\Psi\rangle=-\langle\Phi^4|\Psi\rangle$,
where the $|\Phi^i\rangle$ are the four configurations shown in Fig. 2.
This symmetry corresponds to the $d_{x^2-y^2}$ symmetry, in agreement with
previous theoretical studies of the Hubbard and $t--J$ models
\cite{Da94,CP92}.

The quasiparticle band structure for two holes has also been investigated.
The main interactions $t_i$ and the overlaps $s_i$ 
between the configurations are given in Table VI (the meaning of
the symbols is specified in Fig. \ref{bipolaronCI}).
The results correspond to a $6 \times 6$ cluster with the two holes
at the shortest distance. At finite $U$ many interactions
contribute to the band, in Table VI we only show the largest ones.
Of particular significance is the $t_3$ interactions which accounts
for the simultaneous  hopping of the two holes. This term, which clearly
favors pairing, vanishes in the infinite $U$ limit, in line  
with the results for the hole--hole
interaction (see above) which indicate that pairing is not favored at
large $U$. Also in this limit $t_1=t_2$. 
Including only the interactions given in Table VI, the bands can 
be easily obtained from,
\begin{equation}
\left |
\begin{array}{cc}
E+2(s_1E-t_1){\rm cos}k_x + 2(s_3E-t_3){\rm cos}k_y &  
(s_2E-t_2)(1+{\rm e}^{{\rm i}k_x})(1+{\rm e}^{-{\rm i}k_y})  \\
(s_2E-t_2)(1+{\rm e}^{-{\rm i}k_x})(1+{\rm e}^{{\rm i}k_y})  &
E+2(s_1E-t_1){\rm cos}k_y + 2(s_3E-t_3){\rm cos}k_x   
\end{array}
\right |=0
\end{equation}
Neglecting the overlap, the bands are given by,
\begin{equation}
E({\bf k})=(t_1+t_3)({\rm cos}k_x + {\rm cos}k_y) \pm 
\sqrt{[(t_1+t_3)({\rm cos}k_x + {\rm cos}k_y)]^2+4t_2^2
(1+{\rm cos}k_x)(1+{\rm cos}k_y)}
\end{equation}
In the infinite $U$ limit ($t_3=0$ and $|t_1|=|t_2|$)  the bands are simply,
\begin{mathletters}
\begin{equation}
E_1({\bf k})=-2t_1
\end{equation}
\begin{equation}
E_2({\bf k})=2t_1(1+{\rm cos}k_x+ {\rm cos}k_y)
\end{equation}
\end{mathletters}
\noindent Note that, as in the single hole case and due to the 
antiferromagnetic background, the bands have twice the lattice periodicity.
The dispersionless band has also been reported in \cite{CK98}
and, in our case, it is a consequence of the absence of two hole hopping
in the infinite $U$ limit ($t_3=0$).
Our results, however, disagree with the conclusions reached in \cite{CK98}
concerning the absence of hole attraction.
We find a finite attraction for holes at intermediate $U$'s. It is interesting
to note that our effective hopping is of order $t$, and not of order $t^2/U$
as in cite{CK98}.
This effect is due to the delocalized nature of the single polaron texture
(5 sites, at least), and it does not correspond to a formally similar term
which can be derived from the mapping from the Hubbard to the 
t-J model\cite{Tr88}.

The results for the hole--hole correlation,
$\langle ( 1 - n_i ) ( 1 - n_j ) \rangle$, as function of
the hole--hole distance $r_{ij}=|{\bf r}_i-{\bf r}_j|$ are
reported in Tables VII and VIII. The normalization 
$\sum_j\langle ( 1 - n_i ) ( 1 - n_j ) \rangle = 1$ has been used. The results
correspond to CI wavefunctions with $S_z=0$ and were obtained including
all configurations from either the set \{1,0\} or from the sets
\{1,0\}, \{1,2\}, \{3,0\} and \{2,3\}. The results are in qualitative
agreement with those in \cite{FO90}.  When comparing with
results obtained for the t-J model, one must take into account
that, in the Hubbard model, the hole--hole correlation, as defined
above, can take negative values (see Appendix). This is due to the 
appearance
of configurations with double occupied sites, which are counted as
negative holes. Aside from this effect, our results describe
well a somewhat puzzling result found in the t-J model
(see for instance \cite{GM98,WS97,RD97}):
the maximum hole--hole correlation occurs when the two holes
are in the same sublattice, at a distance equal to $\sqrt{2}$
times the lattice spacingi \cite{diagonal}. 
This result follows directly from the
delocalized nature of the spin polarons, as seen in Fig. \ref{spins}.
The center of each spin polaron propagates through one sublattice only,
but the electron cloud has a finite weight in the other one, 
even when $U/t \rightarrow \infty$. This effect is noted in all 
cases but for $U=200$ with all configurations. In that case there is
not a clear maximum and the correlations are appreciable even at rather
large distances. The reason for this behavior is  that for
large $U$ the configuration with the holes at the largest distance, namely,
\{2,3\}, have the lowest energy and, thus, a large weight in the
CI wavefunction. This is consistent with the fact that no
attraction was observed at large $U$ (see Fig. \ref{difference}).
Finally we note that the slower decrease with distance of hole--hole 
correlations, obtained for $U=8$  including configurations from the
four sets (Table VIII) may be a consequence of the decrease in the 
difference between UHF and CI energies as $U$ diminishes (see Fig. 4).

\subsection{Four Holes}
An interesting question is whether the holes would tend to segregate
when more holes are added to the cluster. In order to investigate this 
point, we have calculated total energies for four holes
on  $10 \times 10$ clusters with the holes either centered on a square, 
or located on two bipolarons separated by a (5,5) vector and with the holes
at the shortest distance. Two (four)
configurations (plus translations) were included in each case.
In the case of two bipolarons only configurations 
in which the two bipolarons are rotated simultaneously are included.
Other possible configurations have different energies and contribute to a less
extent to the wavefunction. In any case, increasing the size of the basis 
set would not have changed the essential conclusion of our analysis (see below).
The results for several values of $U$ are shown in Table IX.
We note that already at the UHF level the solution with two
separated bipolarons has a lower energy. The Coulomb repulsion term in
the Hamiltonian does not favor the configuration with the aggregated
holes but for very small $U$. Restoring lattice symmetry decreases
the energy in both cases to an amount which in neither case vanishes
in the infinite $U$ limit. The decrease is slightly larger in the case
of the four holes on a square. This result can be understood by noting that
the  holes move more freely (producing the smallest distortion to the AF 
background)  when the
charge is confined to the smallest region possible. In any case, this
is not enough to compensate the rather important difference in energy
between the two cases at the UHF level. These results indicate that for large
and intermediate $U$ no hole segregation takes place and
that the most likely configuration is that of separated bipolarons.

\subsection{Effective Hamiltonian for Hole Pairing}
\begin{figure}
\centerline{\epsfig{file=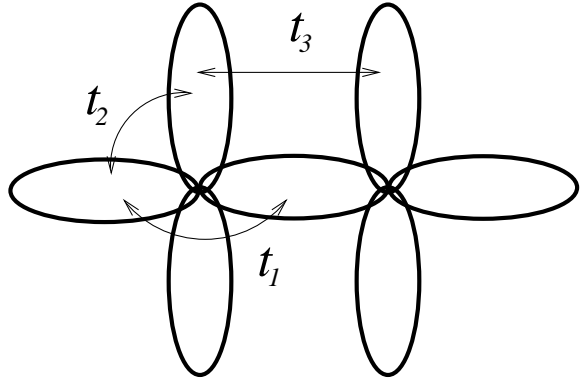,width=3in}}
\caption{
Interactions between two hole configurations having
the holes at the shortest distance (separated by a (1,0) vector).}
\label{bipolaronIN}
\end{figure}

As discussed above, in the large $U$ limit the bipolaron moves over the 
whole cluster due to the interactions among the four mean field 
wavefunctions of Fig. 
\ref{bipolaronCI} (interactions $t_1$ and $t_2$ in Fig. \ref{bipolaronIN}). 
This mechanism can be viewed as another manifestation of hole assisted
hopping.
The possibility of hole assisted hopping has been already considered
in \cite{Hi93}, although in a different context. It always leads 
to superconductivity. In our case, we find a contribution,
in the large $U$ limit, of the type:
\begin{eqnarray}
{\cal H}_{hop} &= &\sum \Delta t {c^{\dag}}_{i,j;s} c_{i,j;s} (
{c^{\dag}}_{i+1,j;{\bar s}} c_{i,j+1;{\bar s}} +  
\nonumber \\ &+ &{c^{\dag}}_{i-1,j;\bar{s}} c_{i+1,j;\bar{s}} + h. c. +{\rm 
perm})
\label{hopping}
\end{eqnarray}
This term admits the BCS decoupling 
$\Delta t \langle c^{\dag}_{i,j;s}
c^{\dag}_{i+1,j;{\bar s}} \rangle c_{i,j;s} c_{i,j+1;{\bar s}} +
h. c.  + ...$. 
It favors superconductivity with
either $s$ or $d$ wave symmetry, depending on the sign of $\Delta t$.
Since we find $\Delta t > 0$, $d$--wave symmetry follows.

\section{Concluding Remarks}
We have analyzed the leading corrections to the Hartree Fock solution
of the Hubbard model, with zero, one and two holes. We show that
a mean field approach gives a reasonable picture of the undoped system
for the entire range of values of $U/t$. 

The main drawback of mean field solutions in doped systems is their
lack of translational invariance. We overcome this problem by using
the Configuration Interaction method. In the case of one hole, the
localized spin polaron is replaced by delocalized wavefunctions
with a well defined dispersion relation. The bandwidth, in the large $U$
limit, scales as $\frac{t^2}{U}$, and the solutions correspond to
spin polarons delocalized in a given sublattice only.
As in the undoped case, these results are in good agreement with
other numerical calculations, for the Hubbard and t-J models.

The same approach is used to analyze the interactions between
pairs of holes. We first obtain Hartree Fock solutions with two holes
at different separations. From comparing their respective energies, and
also with single hole solutions, we find a short range repulsive interaction
between holes. This picture is significantly changed 
when building delocalized solutions.
The energy gain from the delocalization is enough to compensate the
static, mean field, repulsion. There is a net attractive interaction
for $8 \le U/t \le 50$, approximately. The correlations between
the holes which form this bound state are in good agreement with
exact calculations, when available. The state has $d_{x^2-y^2}$
symmetry. In this range of parameters, we find no evidence
of hole clustering into larger structures.

A further proof of the efficiency ot the present CI approach results
from a comparison with the CI approach  of \cite{FL97}, that is 
based upon an extended basis (${\bf k}$--space). For a $6 \times 6$ cluster 
and $U=4$ the UHF localized solution has an energy of -31.747. 
A CI calculation including 36 localized configurations lowers the energy 
down to -31.972. This has to be compared with the result reported in
\cite{FL97} obtained by means of 2027860 extended configurations,
namely, -30.471. The difference between the two approaches should 
further increase for larger $U$.

We have not applied the same technique to other
Hartree Fock solutions which have been found extensively in the
Hubbard model: domain walls separating
antiferromagnetic regions \cite{VL91,ZG89,PR89,Sc90}. 
The breakdown of translational
symmetry associated with these solutions is probably real and not
just an artifact of the Hartree Fock solution, as in the previous cases.
Hybridization of equivalent solutions can, however, stabilize
domain walls with a finite filling, which are not the 
mean field solutions with the lowest energy.

Because of the qualitative differences between spin
polarons and domain walls, we expect a sharp transition between
the two at low values of $U/t$.
Note, however, that the
scheme presented here, based on mean field solutions plus
corrections, is equally valid in both cases.

\acknowledgments
Financial support from the  CICYT, Spain, through grants PB96-0875,
PB96-0085, PB95-0069, is gratefully acknowledged.

\section{Appendix: Hole-hole Correlations in the Hydrogen Molecule}

Here we explicitly calculate the hole--hole correlations in the hydrogen
molecule described by means of the Hubbard model. Let us
call $a^{\dagger}_{\sigma}$ and $b^{\dagger}_{\sigma}$ the operators
that create a particle with spin $\sigma$ at sites $a$ and $b$ 
respectively. The ground state wavefunction has $S_z=0$ and is given by,
\begin{equation}
|\psi> =(2+\alpha^2)^{-1/2}\left(|\phi_1>+|\phi_2>+|\phi_3>\right ) 
\end{equation}
\noindent where,
\begin{mathletters}
\begin{equation}
|\phi_1> = a^{\dagger}_{\uparrow}b^{\dagger}_{\downarrow}|0>
\end{equation}
\begin{equation}
|\phi_2> = \frac{1}{\sqrt{2}}\left(
a^{\dagger}_{\uparrow}a^{\dagger}_{\downarrow}+
b^{\dagger}_{\uparrow}b^{\dagger}_{\downarrow}\right)|0>
\end{equation}
\begin{equation}
|\phi_3> = b^{\dagger}_{\uparrow}a^{\dagger}_{\downarrow}|0>
\end{equation}
\label{H2}
\end{mathletters}
\noindent with
\begin{equation}
\alpha=\frac{E}{\sqrt{2}},\;\;\; 
E=\frac{U}{2}-\left(\frac{U^2}{4}+4\right)^{1/2}
\end{equation}
\noindent The wavefunctions in Eq. (\ref{H2})  are orthonormalized, namely, 
$<\phi_i|\phi_j>= \delta_{ij}$. The result for the hole--hole correlations 
on different sites is,
\begin{equation}
<\psi|(1-n_a)(1-n_b)|\psi>=-\frac{\alpha^2}{2+\alpha^2}
\end{equation}
\noindent As $\alpha=-\sqrt{2},0$ when $U=0,\infty$, this expectation
value varies from -0.5 to 0.0. Thus it can take negative values as found
in the case of clusters of the square lattice. Instead, the  hole--hole 
correlation on the same site is given by,
\begin{equation}
<\psi|(1-n_a)(1-n_a)|\psi>=\frac{\alpha^2}{2+\alpha^2}
\end{equation}
\noindent which is positive for all values of $U$. Particle--particle
correlations are obtained by adding 1 to these results.

\newpage

\begin{table}
\caption{
Exact {\protect \cite{FO90}}, UHF and CI (see text) energies of the 
Hubbard model versus $U$, for the half--filled 4 $\times$ 4 cluster.
All energies are given
in units of the hopping integral $t$. The UHF solution of lowest
energy corresponds to the antiferromagnetic (AF) configuration
while the CI results were obtained by adding to the AF
configuration the 32 configurations having two neigboring spins flipped 
(referred to as sf). The interaction and overlap between
the AF and the sf configurations ($S_{\rm {AF,sf}}$ and $t_{{\rm AF,sf}}$)
are also given.}
\label{Tab_exact}
\begin{tabular}{lcccccc}
    & Exact  & \multicolumn{2}{c} {UHF} & CI & &  \\
\tableline
   U  & Exact  & AF  & sf & AF+sf & $t_{{\rm AF,sf}}$ & $S_{\rm {AF,sf}}$ \\
\tableline
  6 &-10.55222 & -9.37989 & -7.94663 & -9.85296 &  1.1410 &  0.1035 \\
  8 & -8.46887 & -7.38963 & -6.16154 & -7.87777 &  0.5863 &  0.0572 \\
 16 & -4.61186 & -3.91165 & -3.20231 & -4.27171 &  0.1692 &  0.0150 \\
 32 & -2.37589 & -1.98846 & -1.61884 & -2.19194 &  0.0682 &  0.0039 \\
 50 & -1.53078 & -1.27695 & -1.03838 & -1.41060 &  0.0415 &  0.0016 \\
100 &   -      & -0.63962 & -0.51980 & -0.70736 &  0.0202 &  0.0004 \\
\end{tabular}
\end{table}

\begin{table}
\caption{
UHF and CI (see text) energies of the Hubbard model in
the 4 $\times$ 4 cluster for several values of $U$ and 1 and 2
holes (with respect to half--filling). All energies are given
in units of the hopping integral $t$. The CI results for 2 holes 
were obtained by including all configurations having the holes separated 
by vectors of the sets \{1,0\} or \{2,1\}. The CI calculation for 1 hole
includes the spin polaron configuration. In both cases the full basis set was 
constructed by restoring point and translational symmetries.}
\label{Tab_4x4_1}
\begin{tabular}{lcccccc} 
    &\multicolumn{2}{c}   1 & \multicolumn{4}{c}  2    \\
\tableline
  U &  UHF  & CI & UHF-(1,0) & CI-\{1,0\} & UHF-(2,1) & CI-\{2,1\}  \\
\tableline
  4 &-13.30222 &-13.36021  &-14.09307  &-14.20511 &-14.09240 &-14.16640 \\ 
  6 &-10.27549 &-10.59418  &-11.18149  &-11.65669 &-11.12471 &-11.55214 \\
  8 & -8.46151 & -8.76271  & -9.47288  & -9.94496 & -9.47043 & -9.81703 \\
 16 & -5.37431 & -5.53656  & -6.59120  & -7.08701 & -6.79804 & -6.97274  \\
 32 & -3.70563 & -3.78024  & -5.04100  & -5.54390 & -5.40559 & -5.48360  \\
 50 & -3.09365 & -3.13879  & -4.47406  & -4.97558 & -4.90015 & -4.94714  \\
\end{tabular}
\end{table}  

\begin{table}
\caption{
Exact and  UHF or CI (see text) energies of the Hubbard model in
the 4 $\times$ 4 cluster for several values of $U$  1 or 2
holes (with respect to half--filling). All energies are given
in units of the hopping integral $t$. The CI results for 2 holes 
were obtained by including all \{1,0\} and \{2,1\} configurations 
(see also caption of Table II).}
\label{Tab_4x4_2}
\begin{tabular}{lcccc} 
    & \multicolumn{2}{c} {Exact}  &  \multicolumn{2}{c}  {CI} \\
\tableline
  U & 1 & 2 & 1 & 2 \\
\tableline
  4 &-14.66524 &-15.74459  &-13.36021 &-14.48135  \\ 
  6 &-11.96700 &-13.42123  &-10.59418 &-11.81262 \\
  8 &-10.14724 &-11.86883  & -8.76271 &-10.17505 \\
 16 & -6.80729 & -9.06557  & -5.53656 & -7.21100  \\
 32 & -4.93556 & -7.56832  & -3.78024 & -5.59560  \\
 50 & -4.25663 & -7.07718  & -3.13879 & -5.00862  \\
\end{tabular}
\end{table}  

\begin{table}
\caption{
Same as Table II for two holes wavefunctions with $S_z=1$. The energy of the CI
solution corresponding to (2,2) almost coincides  with that of its 
UHF solution. Including all configurations of the sets \{1,1\} and \{2,2\},
does not change the result obtained with only the former set.}
\label{Tab_4x4_3}
\begin{tabular}{lccc} 
    &\multicolumn{2}{c}  {UHF} &  CI    \\
\tableline
  U &  (1,1) & (2,2) & \{1,1\}   \\
\tableline
  6 &-11.12980 &-10.89495 &-11.77834 \\
  8 & -9.48701 & -9.37275 &-10.11391 \\
 16 & -6.77454 & -6.78954 & -7.10740  \\
 32 & -5.33285 & -5.41009 & -5.48456  \\
 50 & -4.80587 & -4.90506 & -4.89772  \\
\end{tabular}
\end{table}  

\begin{table}
\caption{
UHF and CI (see text) energies of the Hubbard model in
the 6 $\times$ 6 cluster for two holes (with respect to half--filling) 
and several values of $U$. All energies are given
in units of the hopping integral $t$. The UHF results correspond to the
configurations with the two holes at the shortest distance with
either $S_z=0$ or 1 separated by lattice vectors (1,0) or (1,1), 
respectively, and
at the largest distance with $S_z=0$ -holes separated by a (2,3) vector.  
The CI results  
were obtained by including all configurations derived either from the sets
\{1,0\} or \{2,3\} (a total of 72 configurations), the set \{1,1\} (36
configurations) all sets having $S_z =0$, namely, \{1,0\}, \{2,1\}, 
\{3,0\} and \{2,3\} (324 configurations) or all
sets having $S_z=1$, i.e., \{1,1\}, \{2,2\}, \{3,1\} and \{3,3\} 
(117 configurations).}
\label{Tab_6x6}
\begin{tabular}{lcccccccc} 
    & \multicolumn{3}{c} {UHF}  & \multicolumn{5}{c} {CI} \\
\tableline
  U & (1,0) & (2,3) & (1,1) & \{1,0\} & \{2,3\}&\{1,1\} & 
all $S_z=0$ &all $S_z=1$ \\
\tableline
  6&-23.147&-23.121&-23.063 & -23.704 &-23.608&-23.352 & -23.813 & -23.700 \\
  8&-18.867&-18.920&-18.848 & -19.488 &-19.385&-19.075 & -19.586 & -19.448 \\
 20&-9.952&-10.260& -10.159 & -10.525 &-10.452&-10.285 & -10.562 & -10.479 \\
 200&-4.120&-4.632&  -4.500 &  -4.622 &-4.647& -4.510 &  -4.652 &  -4.650 \\
\end{tabular}
\end{table}  

\begin{table}
\caption{
Absolute value of the interactions $t_i$ and overlaps $s_i$ between 
the configurations
included in the case of two holes at the shortest distance,
for several values of the interaction parameter $U$. The
meaning of the symbols is given in Fig. \ref{bipolaronIN}. The
results correspond to the $6 \times 6$ cluster.}
\label{Tab_HS}
\begin{tabular}{lcccccc} 
  U &  $t_1$  & $t_2$ & $t_3$ & $s_1$ & $s_2$ & $s_3$ \\
\tableline
  8  &2.459 &4.672 & 1.202 & 0.124  & 0.235 &0.056 \\
 20  &1.418 & 1.624 & 0.159 & 0.119  &0.135 &0.009  \\
200  & 0.646 &0.651 &0.007 &0.088  & 0.088 & 9$\times 10^{-5}$  \\
2000 &0.584 & 0.584 & 7$\times 10^{-4}$ &0.084 & 0.084 &9$\times 10^{-7}$  \\
\end{tabular}
\end{table}  

\begin{table}
\caption{
Hole--hole correlations $\langle (1-n_i)(1-n_j) \rangle$ 
as a function of the
hole--hole distance $r_{ij}$ for three values of $U$. The results 
correspond to two holes in $6 \times 6$ clusters  
and were obtained including all configurations of the set
\{1,0\}. The normalization
$\sum_j\langle ( 1 - n_i ) ( 1 - n_j ) \rangle = 1$ was used.}
\label{Tab_co1}
\begin{tabular}{lccc} 
 $r_{ij}$ / U &8&20&200 \\
\tableline
0&      1.4 &  0.657 & 0.491  \\
1&     -0.171 & -2.51$\times 10^{-3}$ & 3.27$\times 10^{-2}$ \\
$\sqrt{2}$&      3.99$\times 10^{-2}$& 6.30$\times 10^{-2}$ & 
6.97$\times 10^{-2}$ \\
2&      2.0$\times 10^{-3}$ & 1.43$\times 10^{-3}$ & 3.97$\times 10^{-4}$\\
$\sqrt{5}$&      9.6$\times 10^{-3}$ & 9.55$\times 10^{-3}$ &  9.54
$\times 10^{-3}$ \\
$2 \sqrt{2}$&      3.34$\times 10^{-3}$ &  8.43$\times 10^{-4}$ 
&  7.53$\times 10^{-4}$ \\
3&  4.54$\times 10^{-3}$ & 4.18$\times 10^{-3}$ &  6.03$\times 10^{-3}$ \\
$\sqrt{10}$& 2.46$\times 10^{-3}$ & 6.66$\times 10^{-4}$ &  7.53$\times 
10^{-4}$ \\
$\sqrt{13}$&  1.49$\times 10^{-3}$ & 8.51$\times 10^{-4}$ &  
7.53$\times 10^{-4}$ \\
\end{tabular}
\end{table}  

\begin{table}
\caption{
As in Table VII, but including  all configurations from the sets
\{1,0\}, \{2,1\}, \{3,0\} and \{2,3\}.}
\label{Tab_co2}
\begin{tabular}{lccc} 
 $r_{ij}$ / U &8&20&200 \\
\tableline
0&      1.367 &  0.664 & 0.493  \\
1&     -0.183 & -1.08$\times 10^{-2}$ & 5.18$\times 10^{-3}$ \\
$\sqrt{2}$& 2.38$\times 10^{-2}$& 5.18$\times 10^{-2}$ & 7.65$\times 10^{-3}$ \\
2&    4.37$\times 10^{-3}$ & 3.06$\times 10^{-3}$ & 2.36$\times 10^{-2}$\\
$\sqrt{5}$&8.57$\times 10^{-3}$ & 8.99$\times 10^{-3}$ & 1.59$\times 10^{-2}$ \\
$2\sqrt{2}$& 9.48$\times 10^{-3}$& 5.20$\times 10^{-3}$& 7.48$\times 10^{-3}$ \\
3&  1.30$\times 10^{-2}$ & 7.47$\times 10^{-3}$ & 4.43$\times 10^{-2}$ \\
$\sqrt{10}$&8.86$\times 10^{-3}$& 4.64$\times 10^{-3}$ &  2.36$\times10^{-2}$ \\
$\sqrt{13}$& 1.74$\times 10^{-2}$ & 6.4$\times 10^{-3}$&5.06$\times 10^{-3}$ \\
\end{tabular}
\end{table}  

\begin{table}
\caption{
UHF and CI  energies (in units of the hopping integral $t$) for 4 holes in 
10 $\times$ 10 clusters and several values of the interaction $U$.
The results correspond to configurations with either the four holes on 
a square (denominated as 4)
or on two bipolarons (2+2) separated by a (5,5) vector.}
\label{Tab_4h}
\begin{tabular}{lcccc} 
    &\multicolumn{2}{c} {4} & \multicolumn{2}{c}  {2+2}     \\
\tableline
  U & UHF  & CI  & UHF & CI \\
\tableline
   8 & -50.500 & -50.828 & -50.782 & -51.242 \\
  16 & -29.208 & -29.629 & -29.922 & -30.321 \\
  32 & -17.588 & -17.989 & -18.574 & -18.921 \\
 128 &  -8.651 &  -9.005 &  -9.851 & -10.148 \\
 512 &  -6.406 &  -6.744 &  -7.659 &  -7.941 \\
4096 &  -5.750 &  -6.084 &  -7.020 &  -7.297 \\
\end{tabular}
\end{table}  

\begin{references}
\bibitem{ferro}
Note that at large $U$ the ferromagnetic solution sets in.
In Ref. \cite{LC93a} it was shown that the energy of two holes in
a ferromagnetic $L \times L$ cluster could be fitted by
$E/t=-7.942+31.02/L^2$.
Comparing the present numerical results with this energy gives a transition
to the ferromagnetic solution approximately at $U/t=32$, for the cluster
sizes investigated here.
\bibitem{diagonal}
Note that this results is obtained whithout including configurations
which clearly favor it, such as that proposed in \cite{RD97} in which
the two holes lie on neighboring sites along a diagonal and a neighboring
spin is flipped.
\end{references}
\end{document}